\documentclass{ws-p10x7-vpi}
\usepackage{axodraw}

\begin{document}

\title{Constraints on R--parity Violation from Precision Electroweak Measurements}

\author{
Tatsu Takeuchi${}^{(1)}$, 
Oleg Lebedev${}^{(1)}$, and
Will Loinaz${}^{(1,2)}$
}
\address{
${}^{(1)}$IPPAP, Physics Department, Virginia Tech, Blacksburg, VA 24061, USA\\
${}^{(2)}$Department of Physics, Amherst College, Amherst, MA 01002, USA
}

\twocolumn[\maketitle\abstract{
We constrain the size of R--parity violating couplings using precision
electroweak data.
}]

\section{Introduction}

Precision electroweak measurements provide a window to physics beyond
the Standard Model by constraining the size of radiative 
corrections from new particles and interactions.
In this contribution, we summarize the constraints from the LEP and SLD data
on an R--parity violating extension to the MSSM.
Details are presented in Ref.~1.\footnote{For a discussion on 
the constraints on the MSSM itself, see Ref.~2.}

We extend the MSSM with the addition of the
the following terms to the superpotential:
\begin{equation}
  \frac{1}{2} \lambda_{ijk}   \hat{L}_i \hat{L}_j \hat{E}_k 
+             \lambda_{ijk}'  \hat{L}_i \hat{Q}_j \hat{D}_k 
+ \frac{1}{2} \lambda_{ijk}'' \hat{U}_i \hat{D}_j \hat{D}_k\;,
\label{eq:superpotential}
\end{equation}
where $\hat{L}_i$, $\hat{E}_i$, $\hat{Q}_i$, $\hat{U}_i$, and $\hat{D}_i$ are
the MSSM superfields defined in the usual fashion and the subscript $i=1,2,3$ 
is the generation index.  
We focus our attention on these supersymmetric interactions only and
ignore possible R--parity violating soft--breaking terms.\cite{pilaftsis}
This allows us to rotate away any bilinear terms that may 
be present.\cite{hall-suzuki}\footnote{An extension of 
the MSSM with R--parity violation including bilinear and soft--breaking 
terms is discussed in Ref.~5.}

Since the couplings constants $\lambda_{ijk}$, $\lambda'_{ijk}$, and 
$\lambda''_{ijk}$ are arbitrary and do not have any \textit{a priori} 
flavor structure, they generically lead to flavor dependent processes and
corrections to electroweak observables.
In particular, they will give rise to flavor dependent corrections to
the $Zf\bar{f}$ vertices which can be well constrained by 
the $Z$-pole data from LEP and SLD.
Previous works \cite{previous} have already placed bounds on
$\lambda_{ijk}$ of $\mathcal{O}(10^{-2})$ from lepton
universality in low energy charged current processes,
and their effects on $Z$-pole observables are negligible.
$\lambda'_{ijk}$ and $\lambda''_{ijk}$ have been less tightly 
constrained.  However, the simulateneous presence of both terms leads
to unacceptably fast proton decay so we will assume that only one of these
terms is present at a time.

\section{The Corrections}

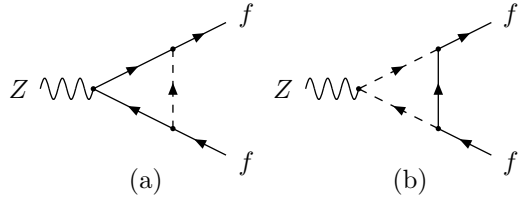
\begin{figure}[tb]
\begin{center} 
\begin{picture}(200,80)(0,-40)
\Vertex(30,0){1}
\Vertex(60,15){1}
\Vertex(60,-15){1}
\Photon(10,0)(30,0){4}{3}
\ArrowLine(30,0)(60,15)
\ArrowLine(60,15)(80,25)
\ArrowLine(80,-25)(60,-15)
\ArrowLine(60,-15)(30,0)
\DashArrowLine(60,-15)(60,15){3}
\Text(2,0)[]{$Z$}
\Text(88,28)[]{$f$}
\Text(88,-28)[]{$f$}
\Vertex(130,0){1}
\Vertex(160,15){1}
\Vertex(160,-15){1}
\Photon(110,0)(130,0){4}{3}
\DashArrowLine(130,0)(160,15){3}
\ArrowLine(160,15)(180,25)
\ArrowLine(180,-25)(160,-15)
\DashArrowLine(160,-15)(130,0){3}
\ArrowLine(160,-15)(160,15)
\Text(102,0)[]{$Z$}
\Text(188,28)[]{$f$}
\Text(188,-28)[]{$f$}
\Text(50,-35)[]{(a)}
\Text(150,-35)[]{(b)}
\end{picture}
\end{center}
\caption{Corrections to the $Zf\bar{f}$ coupling from the 
R--parity violating interactions in Eq.~(\ref{eq:superpotential}).
The sfermions are represented by the dashed lines.
Wavefunction renormalization corrections are not shown.} 
\label{Feynman}
\end{figure}

Corrections to the $Zf\bar{f}$ vertex from the interactions in
Eq.~(\ref{eq:superpotential}) are shown schematically in Fig.~1.
Of all the possible corrections, it can be shown that only those 
with the top quark in the internal fermion line
are numerically significant.
Therefore, we only need to consider
the $\lambda'_{i3k}$ (9 parameters) or the 
$\lambda''_{3jk}$ (3 parameters)\footnote{$\lambda''_{ijk}$ is antisymmetric
in the latter two indices.} interactions.
In this approximation, the $\lambda'_{i3k}$ interactions
only affect the couplings of the left-handed charged leptons and the
right-handed down-type quarks, while
the $\lambda''_{3jk}$ interactions only affect the couplings of
the right-handed down-type quarks.

The actual sizes of these corrections depend on the masses of the
internal sfermions.   As a reference, we choose a common sfermion mass of 
100~GeV.   In this case, the shifts in the $Zf\bar{f}$
couplings due to the $\lambda'_{i3k}$ interactions are found to 
be:\footnote{The tree level coupling of fermion $f$ to
the $Z$ is normalized to $h_f = I_{3f} - Q_f\sin^2\theta_W$.}
\begin{eqnarray*}
\delta h_{e_{iL}}^{\not{R}}
& = & \phantom{-}0.0061 \sum_{k} |\lambda'_{i3k}|^2, \cr
\delta h_{d_{kR}}^{\not{R}}
& = & -0.00215 \sum_{i} |\lambda'_{i3k}|^2.
\end{eqnarray*}
Similarly, the shifts due to the $\lambda''_{3jk}$ interactions are:
\[
\delta h_{d_{jR}}^{\not{R}}
= -0.0043 \sum_{k} |\lambda''_{3jk}|^2.
\]

In order to constrain the size of these shifts, all vertex and
oblique corrections from within the MSSM must also be included and accounted 
for consistently.  Here, we observe that the majority of the $Z$--pole
observables are parity--violating asymmetries or ratios of partial widths
which are all \textit{ratios of coupling constants}.\footnote{The same
observation has been used to constrain a variety of models in Ref.~7.}  
Oblique corrections enter the coupling constants through the $\rho$-parameter 
and the effective value of $\sin^2\theta_W$.\cite{peskin-takeuchi}
The dependence on the $\rho$-parameter cancels in the ratios, 
isolating the effects of oblique corrections in $\sin^2\theta_W$.  
For the vertex corrections, we can apply the
same approximation as above and neglect those without a heavy internal
fermion.  Of the corrections that remain, the simplifying assumption that 
all the sfermion masses are degenerate allows us to either cancel the
correction in the ratios of coupling constants, or absorb them into
a shift in $\sin^2\theta_W$.  The only vertex correction that must be 
considered independently is the Higgs sector induced correction to the
$b_L$ coupling.  These considerations allow us to parametrize all the
corrections from both within and without the MSSM in terms of just a few
parameters which can be fit to the differences of the $Z$-pole data\cite{data}
and ZFITTER\cite{zfitter} predictions.

\section{Lepton Universality}

The shifts in the left-handed couplings of the charged leptons
break lepton universality.  Fitting to the leptonic data
from LEP and SLD, we find:
\begin{eqnarray*}
\delta h_{\mu_L}^{\not{R}} - \delta h_{e_L}^{\not{R}}
& = & \phantom{-}0.00038 \pm 0.00056 \cr
\delta h_{\tau_L}^{\not{R}} - \delta h_{e_L}^{\not{R}}
& = & -0.00013 \pm 0.00061
\end{eqnarray*}
The couplings contributing to $\delta h_{e_L}^{\not{R}}$
are already well constrained from other experiments, so if we neglect them
we obtain the following 1$\sigma$ (2$\sigma$) bounds:
\begin{eqnarray*}
|\lambda'_{23k}| & \le & 0.40\;\;(0.50) \cr
|\lambda'_{33k}| & \le & 0.28\;\;(0.42)
\end{eqnarray*}

\section{Hadronic Observables}

The couplings of the right--handed down--type quarks  
are constrained by the hadronic observables from LEP and SLD.
A global fit to all relevant observables yields:
\begin{eqnarray*}
\delta h_{d_R}^{\not{R}}
& = & 0.081 \pm 0.077 \cr
\delta h_{s_R}^{\not{R}}
& = & 0.055 \pm 0.043 \cr
\delta h_{b_R}^{\not{R}}
& = & 0.026 \pm 0.010
\end{eqnarray*}
Note that $\delta h_{d_R}^{\not{R}}$ and $\delta h_{s_R}^{\not{R}}$
are positive by more than 1$\sigma$, while $\delta h_{b_R}^{\not{R}}$
is positive by more than 2$\sigma$.
Since both the $\lambda'_{i3k}$ and $\lambda''_{3jk}$ interactions
lead to negative shifts, all these couplings are ruled out at the
1$\sigma$ level.  The (2$\sigma$) [3$\sigma$] bounds are:
\begin{eqnarray*}
|\lambda'_{i31}| & \le & (5.8)\;\;[8.4] \cr
|\lambda'_{i32}| & \le & (3.8)\;\;[5.9] \cr
|\lambda'_{i33}| & \le & (\phantom{3.8})\;\;[1.4]
\end{eqnarray*}
or
\begin{eqnarray*}
|\lambda''_{321}| & \le & (2.7)\;\;[4.1] \cr
|\lambda''_{33i}| & \le & (\phantom{2.7})\;\;[0.96]
\end{eqnarray*}
Stronger constrains on $\lambda'_{i31}$ and $\lambda'_{i32}$
are available from other experiments.

\section{Bayesian Limits}

If one makes the \textit{a priori} assumption that the MSSM with
R--parity violation is the correct underlying theory, one obtains
the following 68\% (95\%) Bayesian confidence limits:
\begin{eqnarray*}
\delta h_{d_R}^{\not{R}} & \ge & -0.061\;\;(-0.123) \cr
\delta h_{s_R}^{\not{R}} & \ge & -0.031\;\;(-0.064) \cr
\delta h_{b_R}^{\not{R}} & \ge & -0.0046\;\;(-0.010)
\end{eqnarray*}
This translates into
\begin{eqnarray*}
|\lambda'_{i31}| & \le & 5.2\;\;(7.6) \cr
|\lambda'_{i32}| & \le & 3.8\;\;(5.6) \cr
|\lambda'_{i33}| & \le & 1.4\;\;(2.2)
\end{eqnarray*}
or
\begin{eqnarray*}
|\lambda''_{321}| & \le & 2.7\;\;(3.9) \cr
|\lambda''_{33i}| & \le & 1.0\;\;(1.5)
\end{eqnarray*}
While these Bayesian bounds are considerablely weaker, they
are accompanied by large values of $\chi^2$.

\section{The Common Sfermion Mass}

To obtain bounds for a common sfermion mass other than the
value of $m_{\tilde{f}}=100$~GeV used in this analysis, the limits should
be rescaled by $\sqrt{F(x_0)/F(x)}$, where
\[
F(x) \equiv \frac{x}{1-x}\left( 1 + \frac{1}{1-x}\,\ln\,x\right),
\]
and
\[
x   \equiv \frac{m_t^2}{m_{\tilde{f}}^2},\qquad
x_0 \equiv \frac{m_t^2}{(100\,\mathrm{GeV})^2}.
\]

\section*{Acknowledgments}
Helpful communications with   
B. Allanach, J. E. Brau, R. Clare, H. Dreiner, D. Muller, Y. Nir, 
A. Pilaftsis, F. Rimondi, P. Rowson, D. Su, Z. Sullivan, and M. Swartz
are gratefully acknowledged.
This work was supported in part (O.L. and W.L.) by
the U.~S. Department of Energy, grant DE-FG05-92-ER40709, Task A.


\begin{thebibliography}{99}

\bibitem{rparity}
O. Lebedev, W. Loinaz, and T. Takeuchi,
\textit{Phys. Rev.} D \textbf{61}, 115005 (2000);
\textit{Phys. Rev.} D \textbf{62}, 015003 (2000).

\bibitem{cho}
G.--C. Cho, in this proceedings (hep-ph/0009022).

\bibitem{pilaftsis}
M. Nowakowski and A. Pilaftsis,
\textit{Nucl. Phys.} B \textbf{461}, 19 (1996).

\bibitem{hall-suzuki}
L. J. Hall and M. Suzuki, \textit{Nucl. Phys.} B \textbf{231}, 419 (1984).

\bibitem{kong}
O. C. W. Kong, in this proceedings (hep-ph/0008251).

\bibitem{previous}
B. C. Allanach, A. Dedes, and H. K. Dreiner,
\textit{Phys. Rev.} D \textbf{60}, 075014 (1999).

\bibitem{other}
T. Takeuchi, A. K. Grant, and J. L. Rosner, hep-ph/9409211,
W. Loinaz and T. Takeuchi,
\textit{Phys. Rev.} D \textbf{60}, 015005 (1999),
O. Lebedev, W. Loinaz, and T. Takeuchi,
\textit{Phys. Rev.} D \textbf{62}, 055014 (2000);
hep-ph/0006031,
L. N. Chang, O. Lebedev, W. Loinaz, and T. Takeuchi, hep-ph/0005236.

\bibitem{peskin-takeuchi}
M. E. Peskin and T. Takeuchi,
\textit{Phys. Rev. Lett.} \textbf{65}, 964 (1990); 
\textit{Phys. Rev.} D \textbf{46}, 381 (1992). 

\bibitem{data}
D. Abbaneo, et al., CERN-EP-99-015,
J. Mnich, CERN-EP-99-143,
K. Abe, et al., hep-ex/9908006; hep-ex/9908038.
K. Ackerstaff, et al., \textit{Z. Phys.} C \textbf{76}, 387 (1997),\
E. Boudinov, et al., DELPHI 99-98 CONF 285,
talks by M. Swartz at \textit{Lepton-Photon '99} and
by J. E. Brau, S. Fahey, and G. Quast at \textit{HEP-EPS '99}.


\bibitem{zfitter}
The ZFITTER package: D. Bardin, et al.,
\textit{Z. Phys.} C \textbf{44}, 493 (1989);
\textit{Nucl. Phys.} B \textbf{351}, 1 (1991);
\textit{Phys. Lett.} B \textbf{255}, 290 (1991);
CERN-TH-6443/92;
DESY 99-070, hep-ph/9908433.

\end{thebibliography}
\end{document}